\documentclass{article}

\usepackage{amsmath}
\usepackage{epsf}

\newcommand{\trace}{\mbox{tr}}
\newcommand{\illuseps}[1]{\begin{figure}[ht]\caption{\label{#1}}\centerline{\epsfbox{#1.eps}}\end{figure}}

\begin{document}

\title{Automating First-Principles Phase Diagram Calculations}
\author{A. van de Walle \\
Dept. of Materials Science and Engineering,\\
Northwestern University,\\
Evanston, IL 60208, USA \\
G. Ceder \\
Dept. of Materials Science and Engineering,\\
Massachusetts Institute of Technology,\\
Cambridge, MA 02139, USA}
\maketitle

\begin{abstract}
Devising a computational tool that assesses the thermodynamic stability of
materials is among the most important steps required to build a ``virtual
laboratory'', where materials could be designed from first-principles
without relying on experimental input. Although the formalism that allows
the calculation of solid state phase diagrams from first principles is well
established, its practical implementation remains a tedious process. The
development of a fully automated algorithm to perform such calculations
serves two purposes. First, it will make this powerful tool available to
large number of researchers. Second, it frees the calculation process from
arbitrary parameters, guaranteeing that the results obtained are truly
derived from the underlying first-principles calculations. The proposed
algorithm formalizes the most difficult step of phase diagram calculations,
namely the determination of the ``cluster expansion'', which is a compact
representation of the configurational dependence of the alloy's energy. This
is traditionally achieved by a fit of the unknown interaction parameters of
the cluster expansion to a set of structural energies calculated from
first-principles. We present a formal statistical basis for the selection of
both the interaction parameters to include in the cluster expansion and of
the structures to use in order to determine them. The proposed method relies
on the concepts of cross-validation and variance minimization. An
application to the calculation of the phase diagram of the Si-Ge, CaO-MgO,
Ti-Al, and Cu-Au systems is presented.
\end{abstract}

\section{Introduction}

Steadily growing computer power and improvements in numerical algorithms are
making more and more materials problems approachable by computer
simulations. First-principles computations, in which properties of materials
are derived from quantum mechanics, are particularly interesting as they
allow for the exploration of new materials even before a procedure to
synthesize them has been devised \cite{ceder:ybacuo,vanderven:licoo2}. For
any new material, a determination of its stable structure is of utmost
importance. Hence, the determination of phase diagrams from first principles
is among the most important steps required to build a ``virtual laboratory''.

Over the last twenty years, the formalism enabling the calculation of a
solid-state phase diagram from quantum mechanical energy calculations has
been carefully laid out \cite{fontaine:clusapp,zunger:NATO,ducastelle:book}.
This formalism establishes that the thermodynamic properties of an alloy
can, in principle, be computed as accurately as desired through a technique
known as the \emph{cluster expansion}. In practice, however, the
construction of this expansion can be tedious, and relies on the
researcher's physical intuition to guide the construction process. These
difficulties have so far limited the use of phase diagram calculations from
first-principles to members of the alloy theory community. To make this
powerful tool available to a large number of people outside of the alloy
theory community, we have developed a fully automated algorithm to perform
such calculations and implemented it in an easy-to-use software package. Our
automated algorithm also offers the advantage of removing arbitrary
parameters from the computations, ensuring that the results obtained are
truly derived from the underlying first-principles calculations, rather than
based on the physical intuition of the user.

This paper is organized as follows. First, a brief description of the
formalism to calculate phase diagrams from first principles is presented. We
then propose and motivate an algorithm for the automatic determination of
phase diagrams. Finally, various examples of the application of our method
are shown.

\section{The cluster expansion formalism}

The computation of an alloy phase diagram from first-principles typically
consists of three steps. First, the partition function of the system is 
\textit{coarse grained }to that of a lattice model representing the possible
configurational disorder of the alloy \cite{ceder:ising,avdw:vibrev}. If
needed, this process can be accomplished without neglecting the additional
entropy arising from\ the electronic and vibrational degrees of freedom,
even though these degrees of freedom are no longer explicitly present in the
lattice model. The resulting coarse grained partition function consists of a
sum over every possible way to place the atoms on a given \emph{parent
lattice}.\footnote{%
Here the term ``parent lattice'' is used to distinguish it from the
crystallographic lattice. The parent lattice may have more than one site per
unit cell.} Due to the large number of terms in the sum, it is inconceivable
to attempt to compute the energy of every configuration from first
principles. Hence, in the second step, the dependence of energy on alloy
configuration is parametrized with a simpler model using one of alloy
theory's most powerful tools, the cluster expansion formalism \cite
{sanchez:cexp}. The cluster expansion provides a compact representation of
the configurational dependence of the energy of an alloy, whose accuracy can
be systematically\ improved by adding a sufficient number of terms in the
expansion. In the third step, the system is thermally equilibrated and free
energies are obtained from Monte Carlo simulations. In this paper, we focus
on the cluster expansion, as it is the most difficult step.

The cluster expansion is a generalization of the well-known Ising
Hamiltonian. In the common case of a binary alloy system, the Ising model
consists of assigning a spin-like occupation variable $\sigma _{i}$ to each
site $i$ of the parent lattice, which takes the value $-1$ or $+1$ depending
on the type of atom occupying the site. A particular arrangement of spins of
the parent lattice is called a \emph{configuration} and can be represented
by a vector $\mathbf{\sigma }$ containing the value of the occupation
variable for each site in the parent lattice. Although we focus here on the
case of binary alloys, this framework can be extended to arbitrary
multicomponent alloys (the appropriate formalism is presented in \cite
{sanchez:cexp}, while examples of applications can be found in \cite
{gdg:ternary,mccormack:cd2agau,fontaine:clusapp}).

The cluster expansion parametrizes the energy (per atom) as a polynomial in
the occupation variables: 
\begin{equation}
E(\mathbf{\sigma })=\sum_{\alpha }m_{\alpha }J_{\alpha }\left\langle
\prod_{i\in \alpha '}\sigma _{i}\right\rangle
\end{equation}
where $\alpha $ is a cluster (a set of sites $i$). The sum is taken over all
clusters $\alpha $ that are not equivalent by a symmetry operation of the
space group of the parent lattice, while the average is taken over all
clusters $\alpha '$ that are equivalent to $\alpha $ by symmetry.
The coefficients $J_{\alpha }$ in this expansion embody the information
regarding the energetics of the alloy and are called the effective cluster
interaction (ECI). The \emph{multiplicities} $m_{\alpha }$ indicate the
number of clusters that are equivalent by symmetry to $\alpha $ (divided by
the number of lattice sites).

It can be shown that when \emph{all} clusters $\alpha $ are considered in
the sum, the cluster expansion is able to represent any function $E\left( 
\mathbf{\sigma }\right) $ of configuration $\mathbf{\sigma }$ by an
appropriate selection of the values of $J_{\alpha }$. However, the real
advantage of the cluster expansion is that, in practice, it is found to
converge rapidly. An accuracy that is sufficient for phase diagram
calculations can be achieved by keeping only clusters $\alpha $ that are
relatively compact (\textit{e.g.} short-range pairs or small triplets). The
unknown parameters of the cluster expansion (the ECI) can then determined by
fitting them to the energy of a relatively small number of configurations
obtained, for instance, through first-principles computations. This approach
is known as the Structure Inversion Method (SIM) or the Collony-Williams 
\cite{collwill:fit} method.

The cluster expansion thus presents an extremely concise and practical way
to model the configurational dependence of an alloy's energy. How many ECI
and structures are needed in practice? A typical well-converged cluster
expansion of the energy of an alloy consists of about 10 to 20 ECI and
necessitates the calculation of the energy of around 30 to 50 ordered
structures (see, for instance, \cite
{vanderven:licoo2,gdg:linfit,ozolins:noble}). Once the cluster expansion is
known, various statistical mechanical techniques such as Monte Carlo
simulations \cite{binder:mc}, the low temperature expansion (LTE) \cite
{afk:lte} or the cluster variation method (CVM) \cite
{kikuchi:cvm,ducastelle:book} can be used to obtain phase diagrams.

\section{Optimal Cluster Expansion Construction}

Deciding which ECI are retained in the cluster expansion and which
structural energies are used in the fit is largely a process of trial and
error based on the experience of the researcher, making an automated
procedure difficult. In response to this, we have devised an algorithm that
constructs an optimal cluster expansion by alternatively answering the
following two questions. Given the information obtained at one point in the
calculations regarding the alloy system,

\begin{enumerate}
\item  Which cluster $\alpha $ should be included in the cluster expansion
(nearest neighbor pair, second nearest neighbor pair, triplets, quadruplets,
etc.)?

\item  Which atomic arrangements should be used in order to determine the
unknown coefficients $J_{\alpha }$?
\end{enumerate}

Each question will be treated in turn.

\subsection{ECI selection}

\subsubsection{Cross-validation}

The determination of a cluster expansion differs from standard fitting
procedures by the fact that the number of unknown parameters is
theoretically infinite, since the true physical system cannot be described
exactly with a finite number of nonzero ECI. The problem is that we only
have access to a finite number of structural energies, implying that we can
never determine the exact cluster expansion. The question is then: What is
the best we can do? We obviously need to truncate the series to a finite
number of terms and our focus is to determine the optimal number of terms to
keep, given that a certain number of structural energies are known. If we
keep too few terms, the predicted energies may be imprecise because the
truncated cluster expansion cannot account for all sources of energy
fluctuations. If too many terms are kept, the more insidious problem of
overfitting manifests itself. The mean squared error of the fit may appear
very small, but the true predictive power of the cluster expansion for data
that was not included in the fit is in fact much lower. The source of this
problem is that energy variations caused by one ECI $J_{\beta }$ not
included in the fit may be incorrectly attributed to another ECI $J_{\alpha
} $ which \textit{is} included in the fit, simply because $\left\langle
\prod_{i\in \alpha }\sigma _{i}\right\rangle $ happens to be correlated with 
$\left\langle \prod_{i\in \beta }\sigma _{i}\right\rangle $ in the finite
set of structures whose energies are known. We thus need to find the choice
of ECI that represents the best compromise between those two unwanted
effects. Although it is not widely known, a formal solution to this problem
exists (see for instance, \cite{stone:cv}). It consists in evaluating the
predictive power of a cluster expansion using the cross-validation (CV)
score, defined as: 
\begin{equation*}
(CV)^2=n^{-1}\sum_{i=1}^{n}\left( E_{i}-\hat{E}_{\left( i\right) }\right) ^{2}
\end{equation*}
where $E_{i}$ is the calculated energy of structure $i$, while $\hat{E}%
_{\left( i\right) }$ is the predicted value of the energy of structure $i$
obtained from a least-squares fit to the $\left( n-1\right) $ other
structural energies. Choosing the number of terms that minimizes the CV
score has been shown to be an asymptotically optimal \cite{li:cv} selection
rule. While the formal proof of this result is rather technical, we provide
a heuristic proof of its validity in the Appendix for the convenience of the
reader. In contrast to the well-known mean squared error, the CV score, is
not monotonically decreasing. As the number of parameters to be fitted
increases, the CV score first decreases because an increasing number of
degrees of freedom are available to explain the variations in energy. The CV
score then goes through a minimum before increasing, due to a decrease in
predictive power caused by an increase in the noise in the fitted ECI. The
best compromise between these two effects can then be found.

The idea that a correct measure of the accuracy of the model should involve
attempting to predict data points that are not included in the fit is quite
intuitive and has been employed in previous first-principles phase-diagram
calculations (see, for instance, \cite{ferreira:genstr}). The optimality
property of cross-validation, however, establishes the important fact that
there is no need to partition the data in various ways excluding, 1,2,3,
etc. structures at a time. Simply removing one point at a time is all that
is needed.\footnote{%
Note that this property however relies on the assumption that the errors in
the calculated energies (due, for instance, to numerical noise) are
statistically independent, an assumption that should be justified in the
context of cluster expansion construction.} It is also important to note
that, although the calculation of the CV score involves removing points from
the fit, the CV score nevertheless gives a measure of the predictive power
of the fit obtained with \emph{all} structures included \cite{stone:cv,li:cv}%
. The CV criterion thus enables the assessment of the predictive power of
the cluster expansion while still using all the structural energies
available in order to fit the ECI. There is no need to exclude perfectly
valid data points from the fit when obtaining the ECI.

It would appear that removing one point at a time involves recomputing $n$
distinct regressions, a process exhibiting a computational complexity of
order $n^{2}$. However, the CV score can actually be computed in order $n$
operations, using the formula 
\begin{equation*}
\left( CV\right) ^{2}=n^{-1}\sum_{i=1}^{n}\left( \frac{\left( E_{i}-\hat{E}%
_{i}\right) }{\left( 1-X_{i\cdot }\left( X^{T}X\right) ^{-1}X_{i\cdot
}^{T}\right) }\right) ^{2},
\end{equation*}
which involves only the predicted values $\hat{E}_{i}$ obtained from the
standard ``all points included'' least-squares regression and the matrix $%
X_{i\alpha }$containing the values of $\left\langle \prod_{j\in \alpha
}\sigma _{j}\right\rangle $ for each structure $i$. We let $X_{i\cdot }$
denote row $i$ of the matrix $X$. The order $n$ complexity is achieved by
noting that the matrix $\left( X^{T}X\right) ^{-1}$ can be computed once and
for all.

\subsubsection{Narrowing the Search for ECI}

Applying the cross-validation criterion to the problem of ECI selection
requires a few additional steps. Cross-validation tends to perform better in
practice when the number of choices is kept small, not only because this
minimizes the computational requirements of the search, but also for a more
fundamental reason. The cross-validation score is a statistical quantity and
thus provides an error contaminated measure of the ``true'' quantity we are
interested in. While the statistical noise vanishes in the large sample
limit, it may still affect finite-sample performances. As the number of
alternative cluster choices increases, the likelihood that \emph{one}
suboptimal choice happens to give a smaller cross-validation score than the
true optimal choice increases. By restricting the search only to
``physically meaningful'' candidate cluster expansions, we minimize this
unwanted phenomenon.

Let us now propose a formal definition of what a ``physically meaningful''
cluster choice is. Define the diameter of a cluster as the maximum distance
between two sites in the cluster.

\begin{enumerate}
\item  A cluster can be included only if all its subclusters have already
been included.

\item  An $m$-point cluster can be included only if all $m$-point clusters
of a smaller diameter have already been included.
\end{enumerate}

The first rule formalizes the observation that an $m$-point cluster can be
interpreted as describing the coupling between an $m'$-point
cluster and an $\left( m-m'\right) $-point cluster. Under the
reasonable assumption that coupling terms are less important than each terms
taken separately, we expect a given cluster to be associated with a smaller
ECI than any of its subclusters, implying that subclusters should always be
included in the expansion first. The second rule summarizes the intuition
that large and extended clusters tend to describe weaker interactions than
small and compact ones. By using the diameter as a measure of spatial
extent, it assumes that a cluster is associated with an interaction that is
weaker than the weakest pair it contains, the weakest being presumably the
longest.

While it is possible that the optimal cluster expansion does not satisfy
these requirements, the only consequence of imposing them is that the
algorithm might choose a cluster expansion containing an unnecessarily large
number of terms, with some ECI being set close to zero. These two rules are
sufficient to keep the number of possibilities to a finite and reasonable
number and lead to a particularly simple algorithm to systematically go
through all ``physically meaningful'' cluster choices.

First consider pair clusters in increasing order of length. For each choice
of pair clusters, consider triplets, in increasing order of diameter, up to
the diameter of the longest pair. For each choice of triplets, consider
quadruplets, in increasing order of diameter, up to the diameter of the
largest triplet, etc. Note that when ties between the diameters of different
clusters occur, all clusters of the same diameter are added at the same
time. The fact that the total number of cluster must be less than the number
of structures guarantees that the number of cluster choices to be considered
is finite.

For instance, consider the artificially simple case where the pairs have
length 1,2,3,\ldots and similarly for larger clusters. If one has access to
the energy of 7 structures, the following cluster choices would be
considered (each choice is described by the maximum diameter of the included
pairs, triplets, quadruplets, etc.): (1), (1,1), (1,1,1), (1,1,1,1), (2),
(2,1), (2,1,1), (2,2), (3), (3,1), (4). The number of clusters included is
at most 4, because the empty cluster and the point cluster (assuming there
is only one) are always included, and because cross-validation requires at
least one more structure than there are clusters.

It is possible that for a particular class of systems, more specific rules
can be derived (\textit{e.g. }in fcc-based transition metal alloys, the
fourth nearest neighbor pair is often associated with a larger ECI than the
third nearest neighbor pair \cite{ducastelle:book}), but as a general
system-independent rule, this hierarchy of clusters seems sensible.

\subsubsection{Ground State Prediction}

The accuracy of a cluster expansion is not solely measured by the error in
the predicted energies. It is particularly important that it is able to
predict the correct ground states. Since the mean squared error focuses on
optimizing the absolute energy values, while the ground states are
determined by the ranking of energies, the lowest mean squared error does
not necessarily lead to the most accurate prediction of the ground state
line, as discussed in details in \cite{gdg:linfit}. For this reason, the
search for the best cluster expansion in our algorithm gives an absolute
preference to choices of clusters which yield the same ground states as the
calculated energies. More specifically, a candidate cluster expansion which
has a higher cross-validation score but predicts the correct ground states
will be preferred to a cluster expansion which has a lower cross-validation
score but predicts incorrect ground states.

For a given set of structures, it is possible that no candidate cluster
expansion predicts the right ground states. In these cases, a simple way to
ensure that the ground states are correctly predicted is to give extra
``weight'' to specifically chosen structures in order to obtain the correct
ground states. Although we have already described how to include weights in
the standard least-squares setting, we now have to also define an
appropriate cross-validation score for the purpose of ranking the quality of
weighted least-squares fit. Since a linear regression of $E_{i}$ on $%
X_{i\alpha }$ with weights $w_{i}$ is equivalent to a regression of $%
w_{i}E_{i}$ on $w_{i}X_{i\alpha }$, the natural extension of the
cross-validation score to weighted fits is:

\begin{equation*}
\left( WCV\right) ^{2}=n^{-1}\sum_{i=1}^{n}\left( w_{i}\left( E_{i}-\hat{E}%
_{-i}\right) \right) ^{2},
\end{equation*}
which, for the purpose of calculations, can written as: 
\begin{equation*}
\left( WCV\right) ^{2}=n^{-1}\sum_{i=1}^{n}\left( \frac{w_{i}\left( E_{i}-%
\hat{E}_{i}\right) }{\left( 1-w_{i}X_{i\cdot }\left( X^{T}W^{2}X\right)
^{-1}X_{i\alpha }^{T}w_{i}\right) }\right) ^{2}
\end{equation*}
where $W_{ij}=\delta _{ij}w_{i}$. The following heuristic rule is used to
set the weights. First, assign a unit weight to all structures. If the
ground states are not correctly predicted, flag the ``problematic''
structures as follows. Let $c_{i}$ be the concentration of structure $i$.
Let $g_{j}$ be the indices of the true ground states among the calculated
structures, sorted in increasing order of concentration. If a structure $i$
is such that $\left( \hat{E}_{i},c_{i}\right) $ lies below the line joining $%
\left( \hat{E}_{g_{j}},c_{g_{j}}\right) $ and $\left( \hat{E}%
_{g_{j+1}},c_{g_{j+1}}\right) $ for some $j$, then mark structures $i$, $%
g_{j}$, $g_{j+1}$ as ``problematic''. Increase the weights of all
``problematic'' structures by one and repeat the process until the ground
states are correctly predicted or until a specified number of unsuccessful
trials has been reached. The rationale behind this algorithm is that,
whenever possible, weights should be avoided. By adjusting weights, one can
get a wide variety of results. When only the weights that are strictly
needed to get a qualitatively correct ground state line are adjusted, we
minimize the level of arbitrariness of the fitting procedure.

\subsection{Structure Selection}

How can we select which structural energy to add to an existing fit in order
to improve its accuracy at the least computational cost? Ideally, one would
again select the structure that yields, for a given amount of computational
time, the largest reduction in the prediction error of least-square fit, as
estimated by the CV score. Unfortunately, the CV score requires the
knowledge of the energy of the new candidate structure to be added to the
fit, which makes this approach unfeasible. However, as we will now describe,
it is still possible to construct a useful estimate of the improvement in
the accuracy of a cluster expansion even without the knowledge of the energy
of a candidate new structure to be added to the fit.

\subsubsection{Variance Reduction}

As discussed in more details in the Appendix, the prediction error of a
least-square fit can be separated into two components: the bias and the
variance. These quantities are best described in terms of the following
thought experiment. If you were to choose at random many samples of $n$
structures and\ were to compute a clusters expansion for each sample, each
fit would give slightly different values for the ECI. The ECI, averaged over
different samples, may give a slightly biased estimate of the true value of
the ECI (because a truncated cluster expansion does account for all possible
sources of energy fluctuations). In addition to this systematic bias
component, the ECI fitted to each sample will exhibit fluctuations around
the mean whose magnitude can be characterized by a variance component. Of
course, since the ECI are a multidimensional quantity, a matrix of
covariances is needed to fully characterize the fluctuations.

Ideally, we would want to add to the fit the structure which most reduces
the sum of the bias squared and of the variance. Unfortunately, the
reduction in the bias is impossible to predict without the knowledge of the
energy of the new candidate structure to be added.\footnote{%
One would need to known the value of the ECI obtained with the new structure
included in the fit, which is impossible without the knowledge of its energy.%
} We thus focus solely on the variance component, which can be estimated
without the knowledge of the energy of the new structure.

Is it well known from the theory of least-square estimation \cite
{goldberger:ols} that the covariance matrix of the ECI is given by 
\begin{equation}
V=\left( X^{T}X\right) ^{-1}e^{2}  \label{olsvar}
\end{equation}
where $e^{2}$ is the mean squared error of the fit. Note that $X$, which is
the matrix of the $\left\langle \prod_{j\in \alpha }\sigma _{j}\right\rangle 
$ for each structure, indeed does not depend on the structural energies. (In
principle, the $e^{2}$ term does depend on the structural energy, but it is
a second order effect. To see this, consider $e^{2}$ as a function of the
ECI. Since a least-square fit is obtained by minimizing $e^{2}$, it follows
that $e^{2}$ is a quadratic function of the ECI in the vicinity of the
minimum, implying that a first order change in the estimated ECI yields a
second order change in the estimated $e^{2}$.)

We can now use Equation (\ref{olsvar}) to derive the variance of the
predicted energies. The predicted energy $\hat{E}_{i}$ of structure $i$ is a
linear function of the ECI 
\begin{equation*}
\hat{E}_{i}=\sum_{\alpha }X_{i\alpha }J^*_{\alpha }\equiv X_{i\cdot }J^*_{\cdot }
\end{equation*}
where $J^*_{\cdot}$ denotes the vector of the ECI times their respective
multiplicities (i.e. $J^*_{\alpha}=m_{\alpha} J_{\alpha}$).
The variance of a linear function of a vector $J^*_{\cdot }$ with known
covariance matrix $V$ is given by: 
\begin{equation*}
Var\left[ \hat{E}_{i}\right] =X_{i\cdot }VX_{i\cdot }^{T}.
\end{equation*}
This expression gives the variance of the predicted energy of only \emph{one}
structure. To quantify the predictive power of the cluster expansion, we
then need to average this quantity over all structures (even those not yet
included in the fit). The set of the values of the correlations for every
possible structure is hard to characterize. While it is known to be a convex
polytope embedded in a cube of side 2, constructing this polytope is
computationally intensive \cite{ducastelle:book}. To keep the calculations
tractable, we make the simplifying assumption\ that the correlations $%
X_{i\cdot }$ of every possible structures are distributed isotropically in a
sphere. The expected variance of a structure picked at random is then given
by an integral over a sphere $S$ weighted by some spherically symmetric
density $f\left( \left\| u\right\| \right) $: 
\begin{eqnarray*}
E\left[ Var\left[ \hat{E}_{i}\right] \right] &=&\int_{S}u^{T}Vu\,f\left(
\left\| u\right\| \right) \ du \\
&=&\int_{0}^{\infty }\left( \int_{\left\| u\right\| =r}u^{T}V\ %
u\,du\right) f\left( r\right) dr \\
&=&\left( \int_{\left\| u\right\| =r}u^{T}Vu\,du\right) \int_{0}^{\infty
}f\left( r\right) dr \\
&\propto &\left( \int_{\left\| u\right\| =1}u^{T}Vu\,du\right) \\
&=&\trace\left( V\right) .
\end{eqnarray*}
The last step follows from the fact that for any set of orthogonal vectors $%
u_{i}$, tr$\left( V\right) =\sum_{i}u_{i}^{T}Vu_{i}$. In summary, the trace
of the covariance matrix of the ECI: 
\begin{equation*}
\trace\left( e^{2}\left( X^{T}X\right) ^{-1}\right)
\end{equation*}
provides us with a criterion to estimate the expected variance of the
energies predicted from a cluster expansion.

The question is now: What happens to the expected variance when a new
structure is added to the fit? Changes in $e^{2}$ can be shown to be of a
second order as follows. The least square procedure minimizes $e^{2}$,
implying any first-order change in the ECI (as a result of adding a new
point to the fit) yields no first-order change in $e^{2}$. We can thus focus
solely on finding the new structure $i$ which maximizes the reduction $r$ of
the trace of $\left( X^{T}X\right) ^{-1}$: 
\begin{equation*}
r=\trace\left( \left( X^{T}X\right) ^{-1}\right) -\trace\left( \left(
X^{T}X+X_{i\cdot }^{T}X_{i\cdot }\right) ^{-1}\right)
\end{equation*}
where $X_{i\cdot }$ are the correlations of the new structure to be added.
Since the length of the vector $X_{i\cdot }$ is bounded, there is a limit to
how much a single structure can reduce the variance of the fit. The maximum
variance reduction $\Delta V_{\max }$ is reached when $X_{i\cdot }^{T}$ is
the longest possible column vector $v$ parallel to the eigenvector of $%
X^{T}X $ associated with the smallest eigenvalue: 
\begin{equation}
\Delta V_{\max }=\trace\left( \left( X^{T}X\right) ^{-1}-\left(
X^{T}X+vv^{T}\right) ^{-1}\right)  \label{maxv}
\end{equation}
where $v$ is normalized so that $\max_{j}\left| v_{j}\right| =1$, since
correlations cannot exceed one in magnitude.

This result can be shown by maximizing tr$\left( \left( X^{T}X\right)
^{-1}-\left( X^{T}X+vv^{T}\right) ^{-1}\right) $ with respect to $v$ while
introducing the constraints that $v^{T}v=c$ through a Lagrange multiplier $%
\lambda $. The first-order condition for this maximization problem can be
found by first calculating the first-order change in tr$\left( \left(
X^{T}X\right) ^{-1}-\left( X^{T}X+vv^{T}\right) ^{-1}\right) $ with respect
to changes in $vv^{T}$: 
\begin{eqnarray}
\trace\left( \left( X^{T}X\right) ^{-1}-\left( X^{T}X+vv^{T}\right)
^{-1}\right) &=&\trace\left( \left( X^{T}X\right) ^{-1}-\left( I+\left(
X^{T}X\right) ^{-1}vv^{T}\right) ^{-1}\left( X^{T}X\right) ^{-1}\right)
\label{trlin} \\
&\approx &\trace\left( \left( X^{T}X\right) ^{-1}-\left( I-\left(
X^{T}X\right) ^{-1}vv^{T}\right) \left( X^{T}X\right) ^{-1}\right)  \notag \\
&=&\trace\left( \left( X^{T}X\right) ^{-1}vv^{T}\left( X^{T}X\right)
^{-1}\right)  \notag \\
&=&\trace\left( \left( X^{T}X\right) ^{-2}vv^{T}\right)  \notag \\
&=&\trace\left( v^{T}\left( X^{T}X\right) ^{-2}v\right)  \notag \\
&=&v^{T}\left( X^{T}X\right) ^{-2}v  \notag
\end{eqnarray}
The first-order condition is then given by 
\begin{equation*}
\nabla _{v}\left( v^{T}\left( X^{T}X\right) ^{-2}v-\lambda \left(
v^{T}v-c\right) \right) =0,
\end{equation*}
which is just an eigenvalue problem: 
\begin{equation*}
\left( X^{T}X\right) ^{-2}v=\lambda v.
\end{equation*}
Since we are looking for a maximum, we seek the largest eigenvalue of $%
\left( X^{T}X\right) ^{-2}$ or, equivalently, the smallest eigenvalue of $%
X^{T}X$. This establishes that $v$ must be parallel with the eigenvector
associated with the smallest eigenvalue of $X^{T}X$.

\subsubsection{Trade-off between Computational Requirements and Variance
Reduction}

We now derive an optimality criterion for selecting a new structure to
include in the fit that takes both variance reduction and computational
requirements into account. Consider the change in variance when one
structures is added to a fit already containing $n$ structures. Then $X^{T}X$
is of order $n$, while the change in $X^{T}X$ due to the new structures, $%
X_{i\cdot }^{T}X_{i\cdot }$, is of order $1$. Assuming that $n\gg 1$, a
Taylor series argument similar to the one of Equation (\ref{trlin}) yields: 
\begin{equation*}
\trace\left( \left( X^{T}X\right) ^{-1}\right) -\trace\left( \left(
X^{T}X+X_{i\cdot }^{T}X_{i\cdot }\right) ^{-1}\right) \approx \trace%
\left( \left( X^{T}X\right) ^{-2}X_{i\cdot }^{T}X_{i\cdot }\right) .
\end{equation*}
This shows that when the number $n$ of structures included is large, the
variance-reducing effect of adding a few structures to the fit is linear.
Hence, when faced with the decision to add a structure reducing the variance
by $\Delta V_{1}$ at a computational cost $C_{1}$ or a structure reducing
the variance by $\Delta V_{2}$ at a computational cost $C_{2}$, we should
add the structure which maximizes $\frac{\Delta V_{i}}{C_{i}}$. Indeed, if $%
C_{1}<C_{2}$ one can compute about $\frac{C_{2}}{C_{1}}$ structures similar
to structure $1$ in the time it takes to compute structure $2$. The
resulting variance reduction using structure $2$ is $\Delta V_{2}$, while
the variance reduction using $\frac{C_{2}}{C_{1}}$ structures similar to
structure $1$ is $\frac{C_{2}}{C_{1}}\Delta V_{1}$. Structure $2$ will be
preferable if $\Delta V_{2}>\Delta V_{1}\frac{C_{2}}{C_{1}}$ or 
\begin{equation*}
\frac{\Delta V_{2}}{C_{2}}>\frac{\Delta V_{1}}{C_{1}}.
\end{equation*}
The computational cost of obtaining the energy of a structure can be
estimated in advance from the number of atoms in its unit cell and the known
scaling law of the first-principles method used.\footnote{%
The symmetry of the structure also has some effect on computational costs,
although we neglect this fact to reduce the computational burden of finding
the best structure to add.}

The algorithm to choose the ``best'' structure is thus the following. We
scan through structures in increasing order of computational cost, that is,
in increasing order of unit cell size. This task can be accomplished using
well-known structure enumeration algorithms \cite{ferreira:genstr}. For each
structure, we compute the potential gains for adding it to the fit 
\begin{equation*}
G_{i}=\frac{\trace\left( \left( X^{T}X\right) ^{-1}-\left(
X^{T}X+X_{i\cdot }^{T}X_{i\cdot }\right) ^{-1}\right) }{C_{i}}
\end{equation*}
where $X_{i\cdot }$ and $C_{i}$ are, respectively, the correlations and the
computational cost of structure $i$. The search can be aborted when 
\begin{equation*}
\frac{\Delta V_{\max }}{C_{i}}<G_{i,\max }
\end{equation*}
where $G_{i,\max }$ is the best gain $G_{i}$ found so far and $\Delta
V_{\max}$ is the maximum possible variance reduction, given by Equation (\ref
{maxv}).

\subsubsection{Ground State Prediction}

We have so far solely focused on reducing the variance of the fit. But a
successful cluster expansion must be able to predict the correct ground
states and a special effort must be made to reduce the possibility of
incorrect ground states. To address this issue, whenever a cluster expansion
is constructed, we generate a large number of structures that were not
included in the fit and check whether there are new predicted ground states
among these structures. If there are none, then the minimum variance
criterion for selecting new structures just presented is appropriate,
because the cluster expansion we have is likely to be qualitatively correct
and all we want is to obtain more precise ECI. If new ground states \emph{are%
} predicted, verifying their validity has priority over minimizing the
variance. In this case, the search for the most variance-reducing structure
is thus restricted to the set of newly predicted ground states.

\subsection{Minimal Cluster Expansion}

Before the refinement procedure described above can be used, a ``minimal''
start-up cluster expansion must be constructed and its ECI determined. The
minimal cluster expansion consists of the empty cluster, all the point
clusters and all the nearest-neighbor pairs surrounding all atoms.\footnote{%
Consider a site centered at the origin and sort all other sites (centered at 
$r_{i}$) in increasing order of distance from the origin. The ``nearest
neighbor shell'' can be defined as the sites $r_{1},\ldots ,r_{n}$ where $n$
is the largest $n$ such that $r_{i}\cdot r_{j}<0$ for all $i,j\leq n$. When
there are more than one atom per unit cell, we consider the radius of each
nearest neighbor shell, take the maximum $r_{\max}$, and then take all the
pairs shorter $r_{\max}$.} To determine the ECI, we simply add structures to
the fit in increasing order of computational requirements\footnote{%
Skipping structures whose correlations are coplanar.} until we have one more
structure than there are ECI (so that the cross-validation score can be
calculated). At this point, refinement with the procedures outlined above
can be started.

\section{Implementation}

\subsection{Design goals}

After having described the theoretical principles supporting our automated
algorithm, we are now ready to discuss its practical implementation. While
designing our automated phase diagram calculation system, we sought to
fulfill a number of important requirements.

The system should be able to run over a long period of time without
requiring user intervention while still allowing the user to intervene at
any time, if desired. Such a feature is useful in the context of \textit{ab
initio} calculations because the first few structural energy calculations
are typically set up manually, in order to determine the values of the
various input parameters of the {ab initio} method that offer the best
trade-off between accuracy and computational requirements. Once these
parameters are chosen, they can be kept constant for the subsequent runs and
the remainder of the process can be entirely automated.

The system should keep the user informed of the best result obtained so far,
as the calculation proceeds. In this fashion, the user can interrupt the
process at any time if the currently available results are suitably
accurate. In addition, the user has constant access to preliminary results,
which can be used to try out subsequent processing steps (e.g. Monte Carlo
simulations), while the code keeps running, continuously improving the
precision of the cluster expansion.

Another important requirement is fault-tolerance. As good as modern
first-principles codes are, they sometimes fail and our system should
gracefully handle these instances. The user may also decide at a later time
to recalculate some energies more accurately, and the code should keep track
of such updates.

We wanted our system to be easily adaptable to any computing environment or
energy methods. Many first-principles codes and many computer architectures
are currently available. We cannot simply assume that one code or one system
will eventually replace all others. Finally, we wanted our system to make
maximum use of parallelization opportunities.

\subsection{Organization of the Package}

We achieve the goals described in the previous section through the setup
illustrated in Figure \ref{mapsflow}. The process of fitting a cluster
expansion using the algorithm presented in the previous section is performed
by a subsystem called the MIT Ab-initio Phase Stability (MAPS) code. This
subsystem is completely independent of both the \textit{ab initio} code and
the computing environment used. It implements the algorithms which
determine, at any given time, (i) the best structure to add to a fit and
(ii) the best clusters to add to a fit. The only input this subsystem
requires is the geometry of the lattice, which can typically be described by
a file less than 10 lines long. All other information needed for the
construction of a cluster expansion, such as the space group of the lattice
or a list of candidate clusters and structures, is automatically constructed
from the geometrical input.

The remaining subsystems \emph{are} dependent on the \textit{ab initio} code
and the computing environment used. They consists of various simple
``scripts'' that can be easily adapted to any environment. A first filter
combines a structure's geometric information, generated by MAPS, with
various \textit{ab initio} code-specific input parameters to produce input
files for the first-principles code. In the case of pseudopotential-based
codes, the code-specific parameters include $k$-point mesh density, energy
cutoff and various switches that indicate the type of calculation to be
performed. Less than 10 lines of input are typically needed. Another filter
extracts the structure energies from the output files of the
first-principles calculations.

A ``job manager'' serves as the interface between MAPS and an array of
processors dedicated to structural energy calculations. Communication takes
place through a set of simple signals:

\begin{enumerate}
\item  ``ready'': processors are available to run first-principles
calculations and MAPS should generate new structures whose energy is to be
determined.

\item  ``energy'': one first-principles code has successfully terminated and
the energy of the structure is available.

\item  ``error'': an error has occurred and the energy of a given structure
is not available at this point.
\end{enumerate}

A specific signal is given by simply creating a file of the corresponding
name. In contrast to more sophisticated communication schemes, such a simple
approach is easy to implement and to customize by the end user. It is also
very portable across different computer systems. These files can either be
generated by the user or by simple automated scripts, thus allowing the code
to be either controlled manually or automatically. MAPS simply monitors the
creation of these files and reacts accordingly. If a new structural energy
becomes available (or a new error is reported), a new cluster expansion is
fitted. If a new processor becomes available, a new structure is generated.

Since most structural energy calculations can comfortably run on a single
processor, it is natural to parallelize the process at the level of the
energy calculation. At this level, there is very little need for
inter-process communication, and the parallelization efficiency essentially
reaches 100\%. Any group of workstation can thus be directly set up to
cooperatively construct a cluster expansion without requiring the
installation of sophisticated networks or libraries.

This software package can be obtained by contacting the authors at \texttt{%
avdw@alum.mit.edu}.

\section{Applications}

To illustrate the usefulness of our algorithms and associated software
package, we have computed the phase diagram of a wide variety of alloy
systems: Si-Ge, CaO-MgO, Ti-Al, Cu-Au. Our examples include insulating
(CaO-MgO), semi-conducting (Si-Ge) and metallic systems (Ti-Al and Cu-Au).
The crystal structure of our example systems also exhibits a wide variety,
including systems whose parent lattice has more than one atom per primitive
unit cell (Si-Ge), systems that have spectator ions (O in the CaO-MgO
system), or systems with multiple parent lattices (hcp and fcc in the Ti-Al
system).

With a few exceptions discussed below, our predicted phase diagrams exhibit
the same stable phase as the experimentally determined phase diagram.
However, as is commonly the case in phase diagrams calculated from
first-principles, the temperature scale of the phase transitions is often\
overestimated. The physical origin of this overestimation has not been
unambiguously established, but likely candidates are (i) the omission of the
effect of lattice vibrations \cite{gdg:vib1,gdg:vib2,avdw:ni3al,avdw:vibrev}
(ii) the omission of weak but very long-range elastic interactions due to
atomic size mismatch \cite
{laks:recip,ozolins:nlelas,wolverton:noblesro,ozolins:noble} (iii) the
limited precision of first-principles calculations based on the Local
Density Approximation (LDA). On the one hand, the formalisms that account
for lattice vibrations and elastic interactions have been developed and
would represent a very important addition to our package. On the other hand,
our package can easily be adapted to make use of more accurate
first-principles method, as they become available.

\subsection{Methodology}

Our first-principles calculations are performed within the local density
approximation (LDA) using the VASP \cite{kresse:vasp1,kresse:vasp2} package,
which implements ultra-soft \cite{Vanderbilt:soft_pseudo} pseudopotentials 
\cite{Phillips:pseudo}. In all calculations, we used the default settings
implied by the ``high precision'' option of the code. The $k$-point mesh was
constructed using the Monkhorst-Pack scheme and was chosen such that the
number of $k$-points times the number of atom in the unit cell was at least $%
3500$ for metallic systems and $1000$ for insulating systems. This choice
keeps the $k$-point density constant, despite changes in the unit cell size.
To ensure that the $k$-point density is as isotropic as possible, the number
of mesh points along a given reciprocal lattice vector $a_{1}$ was set
proportional to $\left| a_{1}\cdot \left( a_{2}\times a_{3}\right) \right|
/\left| a_{2}\times a_{3}\right| $, where $a_{2}$ and $a_{3}$ are the two
remaining reciprocal lattice vectors.

Our calculated phase diagrams were obtained through Low Temperature
Expansion (LTE) \cite{afk:lte} and Monte Carlo (MC) simulations \cite
{binder:mc}. The MC simulation cells used ranged from $10\times 10\times 10$
supercells to a $20\times 20\times 20$ supercells. The number of
equilibration MC passes and the number of averaging MC passes used to obtain
thermodynamic quantities were chosen so that the precision of the average
concentration of the alloy is less than $0.1\%$. This precision typically
required from $2000$ to $50000$ equilibration MC passes and from $2000$ up
to $100000$ averaging MC passes, depending on temperature.

The thermodynamic integration method was used to determine the grand
canonical potential $\phi ^{\alpha }$ of each phase $\alpha $ as a function
of chemical potential $\mu $. The grand canonical potential at the starting
point of the integration path was obtained from the LTE. Phase boundaries
between two phases $\alpha $ and $\beta $ are found by locating the
intersection of the curves $\phi ^{\alpha }\left( \mu \right) $ and $\phi
^{\beta }\left( \mu \right) $. The concentrations of each phase in
equilibrium at the phase transition are given by the slope of the curves of
the point of intersection. Additional details regarding these calculations
will be presented in a separate communication \cite{avdw:emc2}.

\subsection{Examples}

While a summary of the characteristics of the five cluster expansions we
constructed can be found in Table \ref{sumce}, the details specific to each
system are discussed below.

\subsubsection{The Si-Ge system}

First-principles investigation of phase stability in the Si-Ge system have
been performed earlier \cite{qteish:sige,gironcoli:sige}. One of the most
thorough studies \cite{gironcoli:sige} relied on the so-called computational
alchemy method. While this ingenious method enables the direct determination
of long-range pair interactions, clusters larger than pairs are not included
in their full generality. (Computational alchemy does allow the pair
interactions to depend on concentration, thereby implicitly accounting, to a
limited extent, for multiplet interactions.) It is thus of interest to
verify that such a simplification is appropriate by employing our automated
tool, which allows the inclusion of multiplets. Figure \ref{sige_eci} shows
that the energetic contribution of pairs indeed dominates the contribution
of multiplets. Not only are the included multiplets associated with very
small ECI, but the cross-validation algorithm automatically determines that
very few triplets should be included in the cluster expansion. Figure \ref
{sige} show the calculated phase diagram. The predicted critical temperature
(325 K) lies between the predictions of \cite{qteish:sige} (375K) and \cite
{gironcoli:sige} (170K).

\subsubsection{The CaO-MgO pseudo-binary system}

A very complete first-principles analysis of the CaO-MgO system was
undertaken in \cite{tepesch:caomgo}. While a substantial amount of labor was
needed to complete this analysis, our software package enabled us to
reproduce this study\footnote{%
Except for the inclusion of lattice vibrations.} in a fully automatic
fashion. In contrast to the Si-Ge system, multiplet interactions are
important in the CaO-MgO system (as seen in Figure \ref{caomgo_eci}) and are
responsible for the pronounced asymmetry in the miscibility gap observed
both in our calculation and in the experimental measurements (see Figure \ref
{caomgo}). This feature was not given as an input to the code and it is
interesting that the code was able to identify it independently.

\subsubsection{The Ti-Al system}

Our third example seeks to reproduce the very detailed first-principles
calculation of the Ti-Al phase diagram found in\ \cite{asta:tial}. We focus
on the Ti-rich portion of the phase diagram because the Al-rich half of the
phase diagram exhibits order-order transitions that would prompt the
inclusion of the thermodynamic effects of lattice vibrations, which is
beyond the scope of the present study. For the same reason, we do not
investigate the Ti (hcp) $\rightarrow $ Ti (bcc) transition.

However, we do model both hcp-based phases (the Ti solid solution and the DO$%
_{19}$ Ti$_{3}$Al phase) and the fcc-based L1$_{0}$ NiAl phase. As shown in
Figure \ref{tial_eci},\ the cluster expansion on the hcp lattice could
easily be pushed to a higher accuracy than in the earlier study \cite
{asta:tial}, which only included up to second-nearest neighbor pair
interactions and a few multiplets. The cluster expansion on the fcc lattice
(see Figure \ref{tialfcc_eci}) does not need to be as accurate because the
location of the phase boundaries was found to be insensitive to further
improvement in the fcc cluster expansion.

Figure \ref{tial} shows our calculated phase diagram. As discussed at the
beginning of this section, our calculations overestimate the experimental
order-disorder transition temperature of the DO$_{19}$ Ti$_{3}$Al phase\ ($%
1450$K). However, our calculated order-disorder transition temperature of $%
1850$K corroborates the first-principles prediction of $1800$K found by \cite
{asta:tial}.

\subsubsection{The Cu-Au system}

The most accurate first-principles study of the Cu-Au system is described in
a series of articles \cite{ozolins:noble,wolverton:noblesro,ozolins:cuau}.
Although our calculations include neither explicit elastic interactions nor
the contribution of lattice vibrations, we were able to obtain good
quantitative agreement with these earlier calculations as well as
experimental measurements, as seen by the transition temperatures reported
in Table \ref{cuautc}. We focus on the Cu-rich half of the phase diagram,
because there is well-documented disagreement between the experimentally
observed ground state at the CuAu$_{3}$ composition and the one predicted
from LDA calculations \cite{ozolins:noble}. As in earlier LDA calculations,
we were not able to model the CuAu-I $\rightarrow $ CuAu-II transition,
which is driven by very small free energy differences that are beyond the
accuracy of current LDA calculations \cite{ozolins:noble}. Our calculated
ECI are plotted in Figure \ref{cuau_eci}, while the resulting phase diagram
is shown in Figure \ref{cuau}.

\section{Conclusion}

We have formalized the decision rules that allow the automatic construction
of a cluster expansion. The development of a fully automated procedure to
compute phase diagrams from first principles allows the powerful cluster
expansion formalism to be used by anyone who possesses basic material
science knowledge. Through a variety of examples, we have shown that earlier
first-principles phase diagram calculations, which represented significant
contribution to the field of alloy theory at the time of their publication,
can now be reproduced in an automated fashion.

A variety of extensions to\ our work would prove extremely useful. A module
that computes the vibrational free energies at a reasonable computational
cost (for instance, through the methods presented in \cite{moruzzi:debyeg}
or \cite{avdw:pd3v,avdw:vibrev}) would allow the modelization of allotropic
transitions and improve the precisions of the calculated transition
temperatures. The inclusion of electronic entropy should be straightforward 
\cite{wolverton:selec}. The explicit inclusion of elastic contributions to
the energy through a reciprocal space cluster expansion \cite
{laks:recip,ozolins:noble} would also be very helpful.

\section*{Acknowledgments}

This work was supported by the U.S. Department of Energy, Office of Basic
Energy Sciences, under contract no. DE-F502-96ER 45571. Gerbrand Ceder
acknowledges support of Union Mini\`{e}re through a Faculty Development
Chair. Axel van de Walle acknowledges support of the National Science
Foundation under program DMR-0080766 during his stay at Northwestern
University.

\appendix

\section{Cross-validation}

\label{appcv}This section provides a heuristic proof of the validity of the
cross-validation selection rule. A formal proof, under the assumptions
relevant to our application, can be found in \cite{li:cv}. The concepts used
are described in more detail in\ \cite{stone:cv,goldberger:ols}.

Let us first describe the conventions underlying our discussion. For a given
choice of clusters $\alpha _{1},\ldots ,\alpha _{k}$, consider a sample of $%
n $ correlation vectors $X_{i\cdot }=\left( X_{i\alpha _{1}},\ldots
,X_{i\alpha _{k}}\right) $, for $i=1,\ldots ,n$. Each correlation vector $%
X_{i\cdot }$ is associated with a large set of structures that have the same
correlation vector $X_{i\cdot }$ (although their other correlations, not
included in $X_{i\cdot }$, differ). We then consider the structural energy $%
E_{i}$ obtained for a structure with correlation $X_{i\cdot }$ to be a
random variable, because fixing $X_{i\cdot }$ does not entirely determine $%
E_{i}$. To handle this randomness, we will be concerned with calculating
expectation values, denoted $\left\langle \cdot \right\rangle $, averaged
over every possible sample of $n$ structures having the given correlation
vectors $X_{i\cdot }$, $i=1,\ldots ,n$. For instance, $\left\langle
E_{i}\right\rangle $ is the\ average energy of all structures with
correlation $X_{i\cdot }$, which is equal to the energy that would be
obtained from a cluster expansion that uses the true exact ECI for all
included clusters $\alpha _{1},\ldots ,\alpha _{k}$ but a zero ECI for all
the other clusters.

Let us now define two important quantities, adopting the convention that $%
i=0 $ denotes a structure (with correlation $X_{0\cdot }$) that is not
included in the least-squares fit.

\begin{itemize}
\item  Let $\hat{E}_{i}$ (for $i=0,\ldots ,n$) be the energy of structure $i$
predicted using a cluster expansion fitted to the $n$ structural energies
available.

\item  Let $\hat{E}_{\left( i\right) }$ (for $i=1,\ldots ,n$) be the energy
of structure $i$ predicted using a cluster expansion fitted to all known
structural energies except the one of structure $i$, for a total of $n-1$
structures.
\end{itemize}

We are now ready to compute the predictive power $P$ of the cluster
expansion, that is, the expected squared error of the prediction of a
structural energy not included in the fit: 
\begin{equation*}
P=\left\langle \left( \hat{E}_{0}-E_{0}\right) ^{2}\right\rangle .
\end{equation*}
In general, for any structure $i$, the expected squared error can be
decomposed as 
\begin{equation}
\left\langle \left( \hat{E}_{i}-E_{i}\right) ^{2}\right\rangle =\left\langle
\left( E_{i}-\left\langle E_{i}\right\rangle \right) ^{2}\right\rangle
+\left\langle \left( \hat{E}_{i}-\left\langle E_{i}\right\rangle \right)
^{2}\right\rangle -2\left\langle \left( E_{i}-\left\langle
E_{i}\right\rangle \right) \left( \hat{E}_{i}-\left\langle
E_{i}\right\rangle \right) \right\rangle ,  \label{biasvar}
\end{equation}
where $\left\langle \left( E_{i}-\left\langle E_{i}\right\rangle \right)
^{2}\right\rangle $ is the so-called bias term while $\left\langle \left( 
\hat{E}_{i}-\left\langle E_{i}\right\rangle \right) ^{2}\right\rangle $ is
the so-called variance term. The covariance term $\left\langle \left(
E_{i}-\left\langle E_{i}\right\rangle \right) \left( \hat{E}%
_{i}-\left\langle E_{i}\right\rangle \right) \right\rangle $ vanishes for a
structure not included in the fit ($i=0$) because $E_{0}$ has no effect on $%
\hat{E}_{0}$. However, for a structure included in the fit ($i>0$), this
covariance term is positive. Whenever $E_{i}$ is large, $\hat{E}_{i}$ will
tend to be large as well, because the least-square procedure attempts to
minimize the distance between $\hat{E}_{i}$ and $E_{i}$. This effect worsens
as the ratio $k/n$ increases. For this reason, attempting to estimate the
predictive power using the mean squared error of the least-squares fit leads
to a biased estimate of the true predictive power. This problem is corrected
by using the cross-validation score, which focuses on the quantity $%
\left\langle \left( \hat{E}_{\left( i\right) }-E_{i}\right)
^{2}\right\rangle $ instead of $\left\langle \left( \hat{E}_{i}-E_{i}\right)
^{2}\right\rangle $. In Equation (\ref{biasvar}), the expected value of the
bias term is unaffected by this change, while the variance term is changed
by an amount that goes to zero as sample size $n$ grows. More importantly,
the covariance term $\left\langle \left( E_{i}-\left\langle
E_{i}\right\rangle \right) \left( \hat{E}_{\left( i\right) }-\left\langle
E_{i}\right\rangle \right) \right\rangle $ now vanishes for $i=1,\ldots ,n$
because the value of $\hat{E}_{\left( i\right) }$ does not depend on $E_{i}$%
. It follows that $\left\langle \left( \hat{E}_{\left( i\right)
}-E_{i}\right) ^{2}\right\rangle $ is an unbiased measure of the predictive
power. By the law of large numbers, the abstract quantity $P=\left\langle
\left( \hat{E}_{\left( i\right) }-E_{i}\right) ^{2}\right\rangle $ can be
estimated by the corresponding sample average $n^{-1}\sum_{i=1}^{n}%
\left( \hat{E}_{\left( i\right) }-E_{i}\right) ^{2}$, the cross-validation
score.

\newpage

\clearpage
\section*{Figures}

%EndExpansion

\illuseps{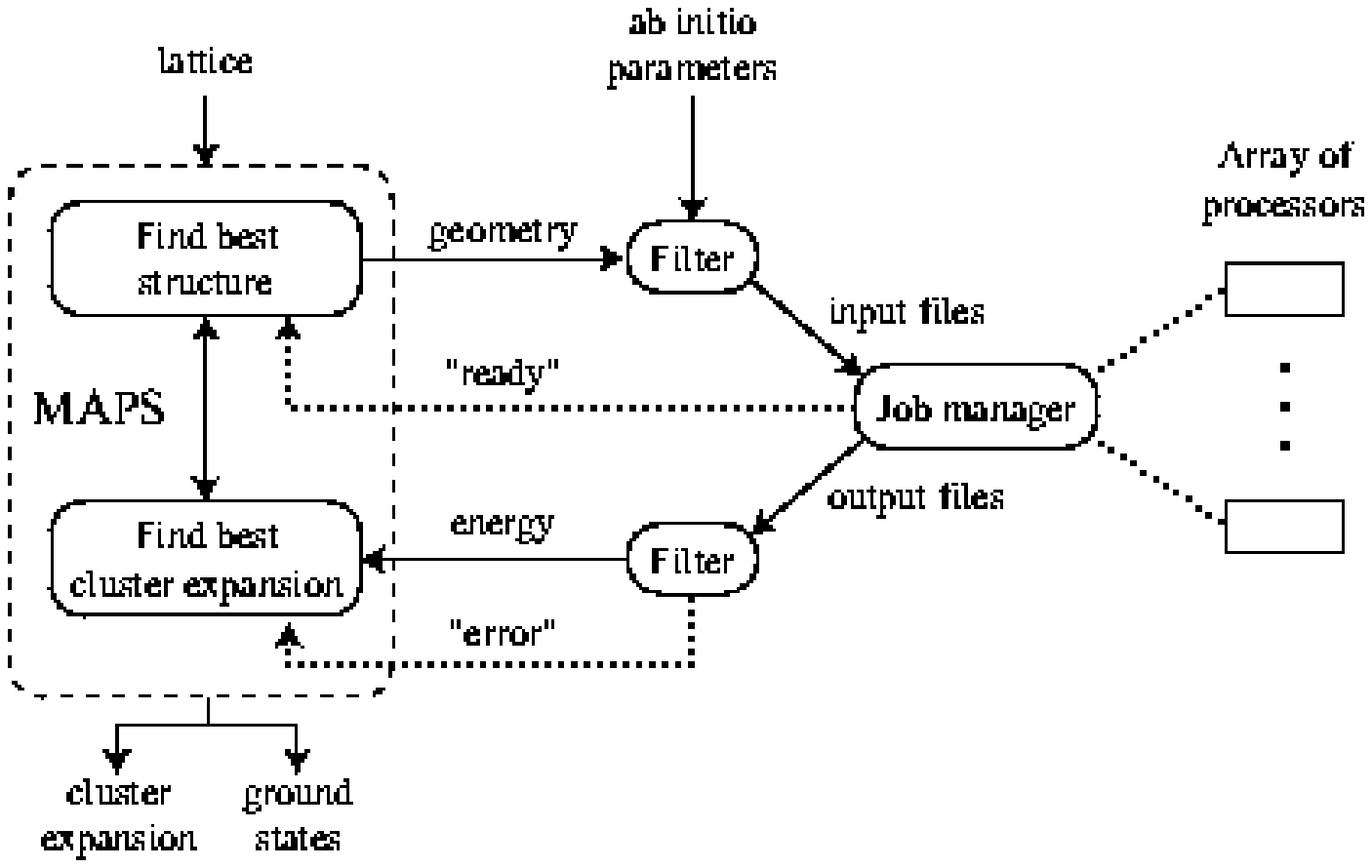}

\illuseps{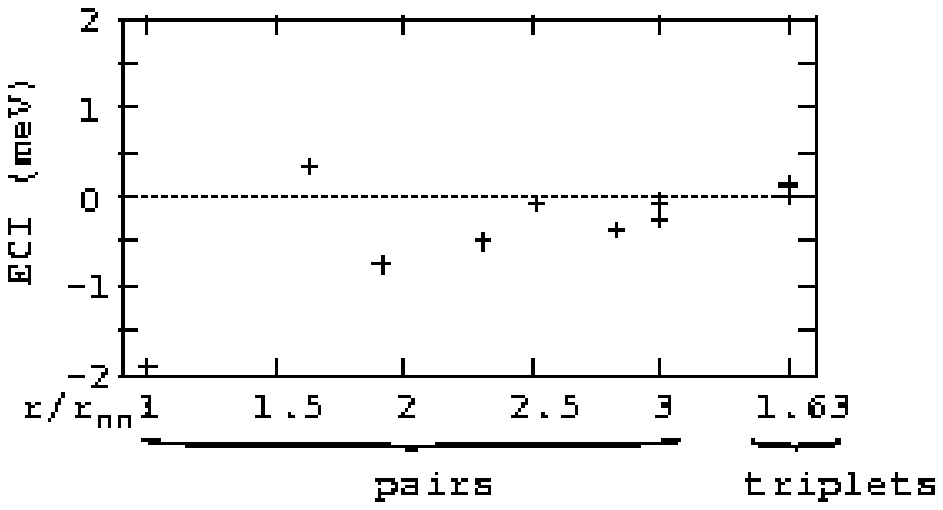}

\illuseps{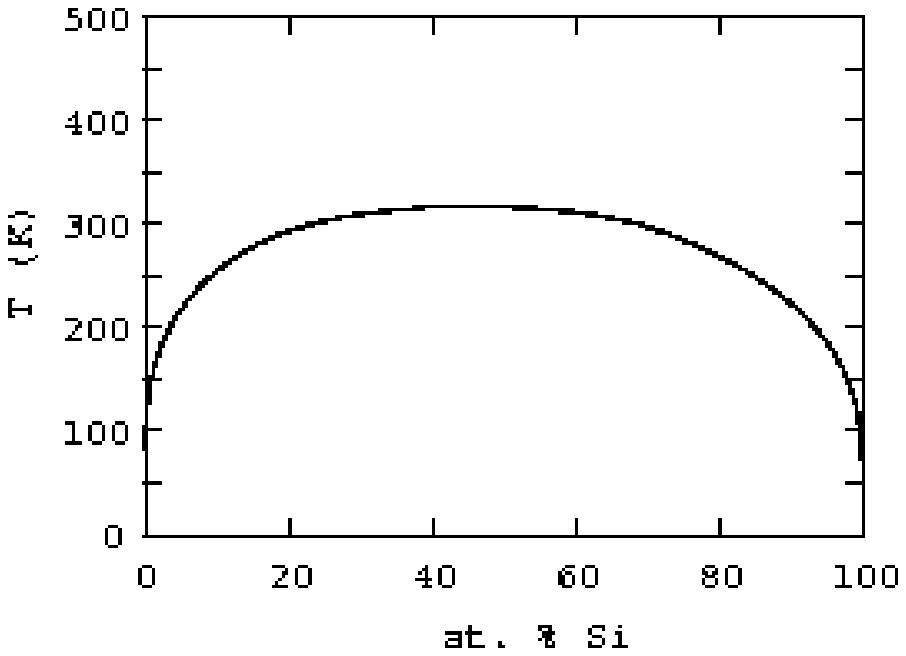}

\illuseps{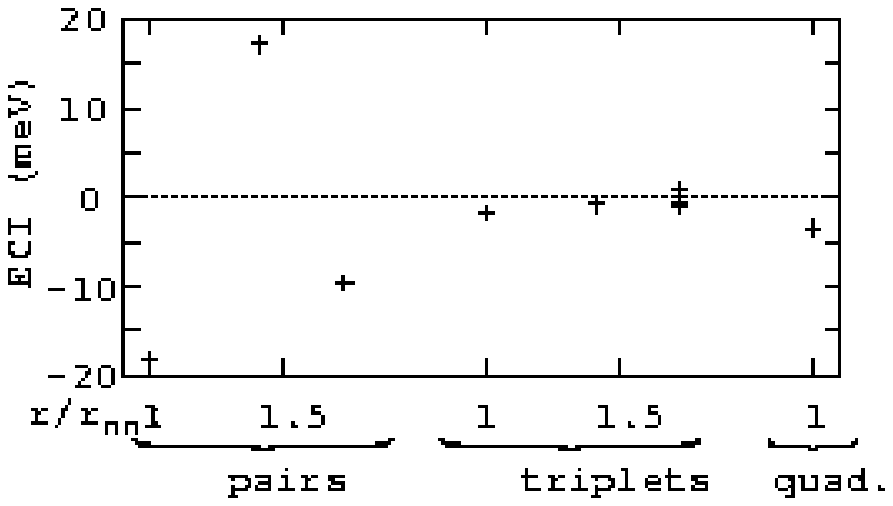}

\illuseps{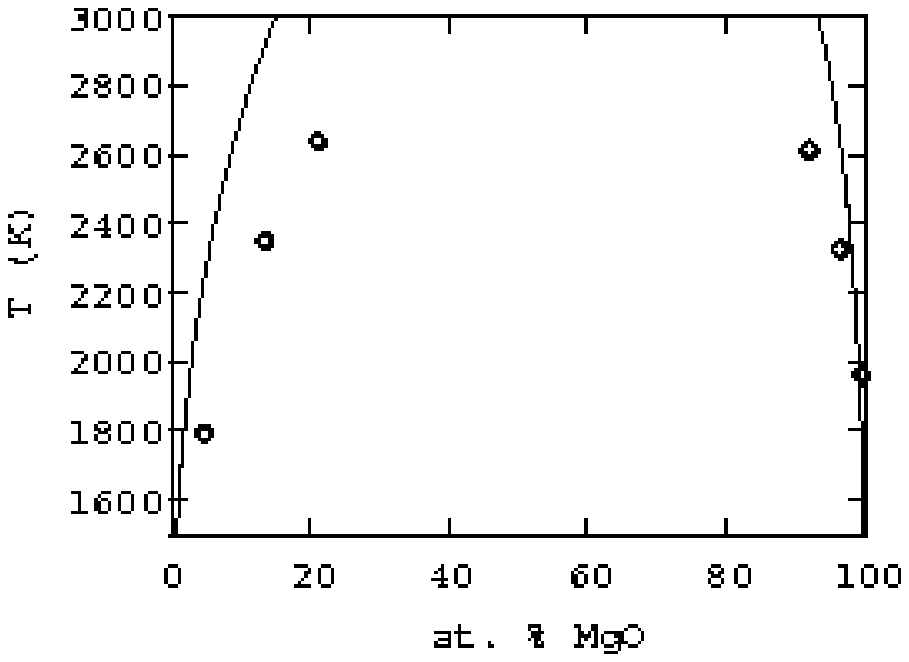}

\illuseps{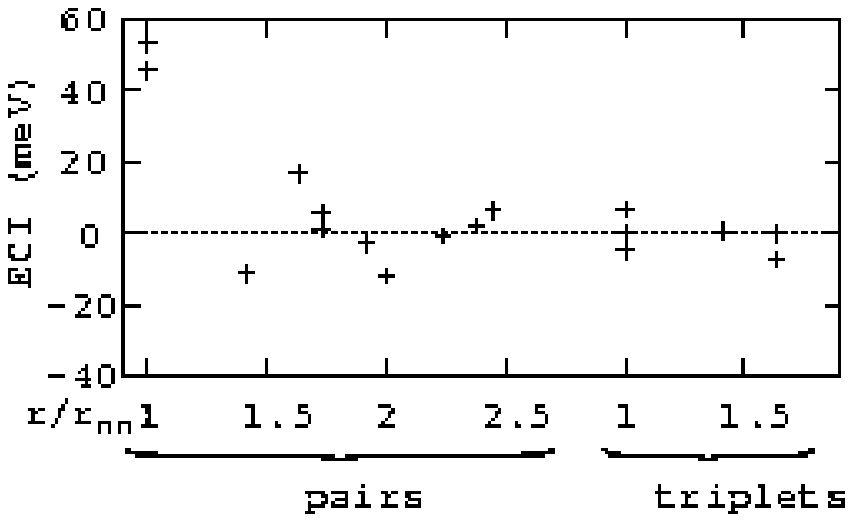}

\illuseps{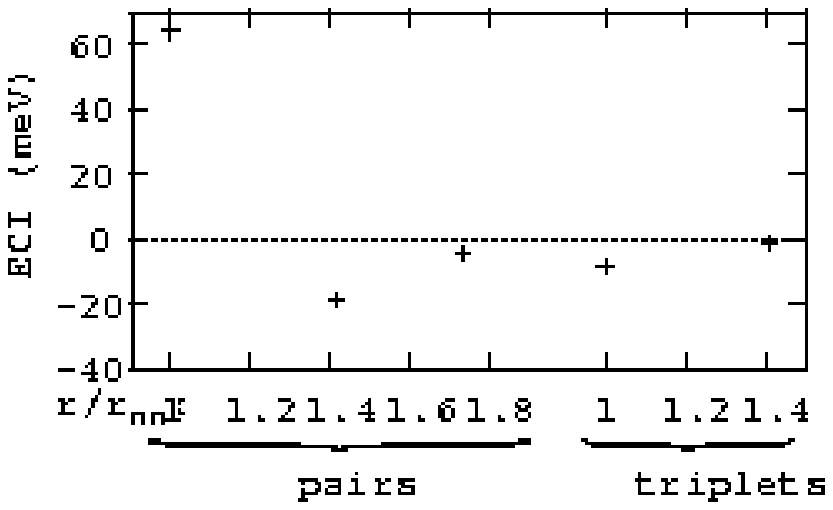}

\illuseps{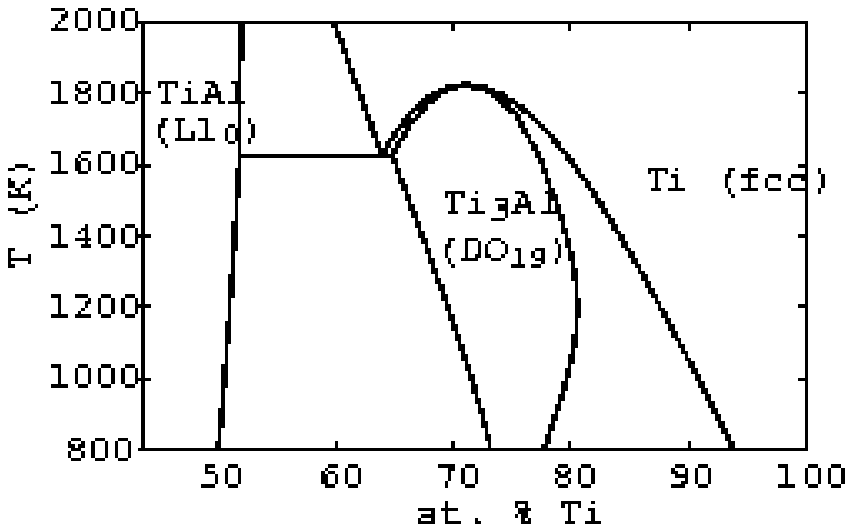}

\illuseps{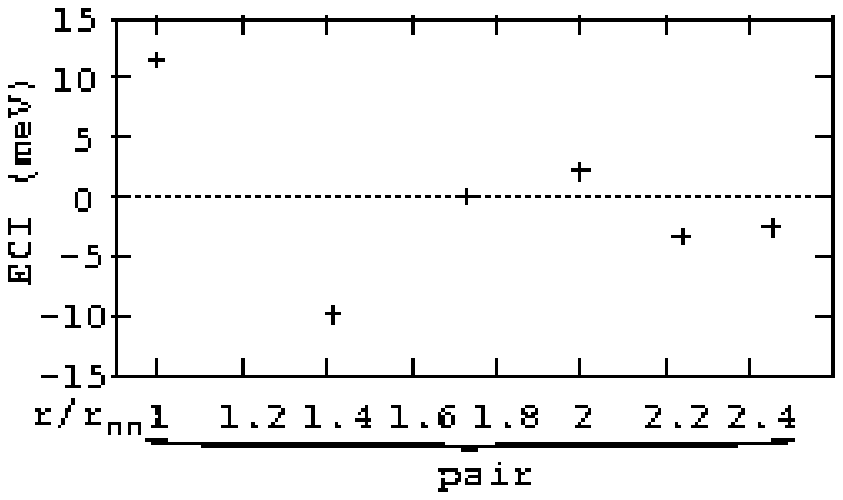}

\illuseps{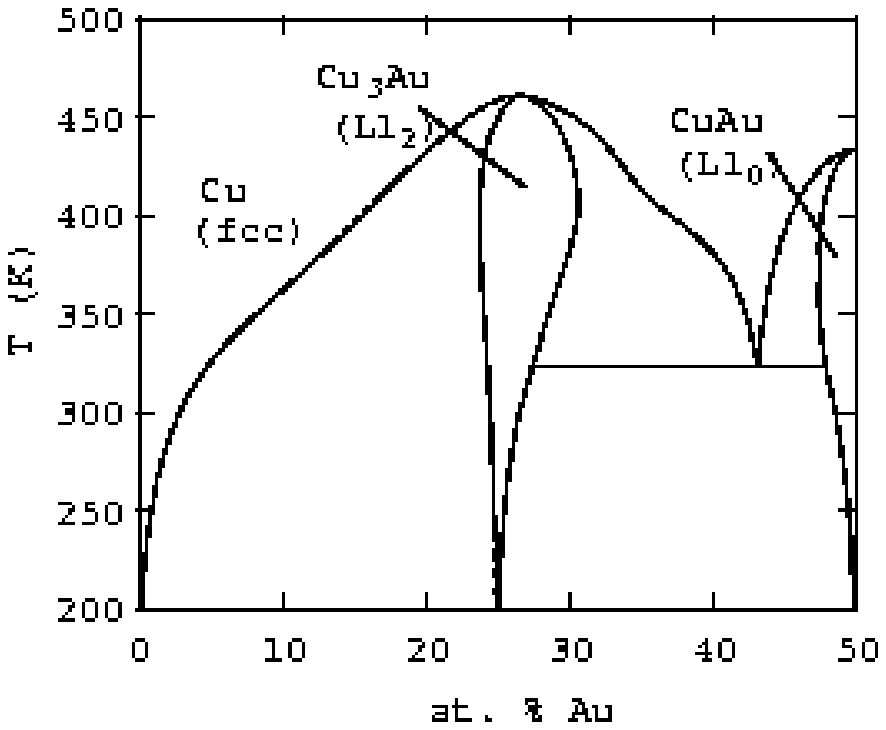}

\clearpage
\section*{Figures Captions}

\noindent Figure \ref{mapsflow}: Organization of the automated cluster
expansion package.

\noindent Figure \ref{sige_eci}: Calculated ECI as a Function of Cluster
Diameter for the Si-Ge System.

\noindent Figure \ref{sige}: Calculated Phase Diagram of Si-Ge System.

\noindent Figure \ref{caomgo_eci}: Calculated ECI as a Function of Cluster
Diameter for the CaO-MgO System.

\noindent Figure \ref{caomgo}: Calculated Phase Diagram of CaO-MgO System.

\noindent Figure \ref{tial_eci}: Calculated ECI as a Function of Cluster
Diameter for the hcp phases of the Ti-Al System.

\noindent Figure \ref{tialfcc_eci}: Calculated ECI as a Function of Cluster
Diameter for the fcc phases of the Ti-Al System.

\noindent Figure \ref{tial}: Calculated Phase Diagram of Ti-Al System
(Ti-rich portion).

\noindent Figure \ref{cuau_eci}: Calculated ECI as a Function of Cluster
Diameter for the Cu-Au System.

\noindent Figure \ref{cuau}: Calculated Phase Diagram of Cu-Au System
(Cu-rich portion).

\clearpage
\section*{Tables}

%TCIMACRO{
%\TeXButton{B}{\begin{table}[h] \centering%
%}}%
%BeginExpansion
\begin{table}[h] \centering%
%
%EndExpansion
\begin{tabular}{llllll}
Characteristic & Si-Ge & CaO-MgO & Ti-Al (hcp) & Ti-Al (fcc) & CuAu \\ 
Number of structures: & \multicolumn{1}{r}{27} & \multicolumn{1}{r}{20} & 
\multicolumn{1}{r}{55} & \multicolumn{1}{r}{23} & \multicolumn{1}{r}{33} \\ 
Number of clusters: & \multicolumn{1}{r}{2+8+3} & \multicolumn{1}{r}{2+3+7+1}
& \multicolumn{1}{r}{2+11+6} & \multicolumn{1}{r}{2+3+2} & 
\multicolumn{1}{r}{2+6} \\ 
CV score (meV/atom): & \multicolumn{1}{r}{1} & \multicolumn{1}{r}{18} & 
\multicolumn{1}{r}{35} & \multicolumn{1}{r}{49} & \multicolumn{1}{r}{23}
\end{tabular}
\caption{Characteristics of the Calculated Cluster Expansions.
The number of clusters is given as the number of each type of multiplet: empty and point clusters+pairs+triplets+quadruplets.\label{sumce}}
%TCIMACRO{
%\TeXButton{E}{\end{table}%
%}}%
%BeginExpansion
\end{table}%
%
%EndExpansion

%TCIMACRO{
%\TeXButton{B}{\begin{table}[h] \centering%
%}}%
%BeginExpansion
\begin{table}[h] \centering%
%
%EndExpansion
\begin{tabular}{llll}
Transition & Present work & Ref. \cite{ozolins:cuau} & Expt. \\ 
Cu$_{3}$Au: L1$_{2}$ $\rightarrow $\thinspace fcc & $460$K & $455$K & $663$K
\\ 
CuAu: L1$_{0}$\thinspace $\rightarrow $\thinspace fcc & $430$K & $560$K & $%
683$K
\end{tabular}
\caption{Calculated and Experimental Transition Temperatures of the Cu-Au
System.\label{cuautc}}%
%TCIMACRO{
%\TeXButton{E}{\end{table}%
%}}%
%BeginExpansion
\end{table}%
%
%EndExpansion

\end{document}